# Goodhart, Charles A.E. and Tsomocos Dimitrios P.: "The Challenge of Financial Stability: A New Model and its Applications"

## Jean-Bernard Chatelain[1]

1st April 2013 (pre-print, forthcoming *Journal of Economics*)

**Abstract:** This review of the book *"The Challenge of Financial Stability: A New Model and its Applications"*[2] by Goodhart C.A.E. and Tsomocos D.P. highlights the potential of the framework of strategic partial default of banks with credit chain on the interbank market for further theoretical and applied research on financial stability.

Since the current financial crisis started in 2007, financial stability is a topic which is at the top of the agenda in economics, but Goodhart and Tsomocos long standing research project on financial stability started much earlier. A number of economists and policy makers now believe that Central Banks should include the objective of financial stability besides inflation targeting. As well, some of them may consider that financial macroeconomics should return to the Keynesian paradigm (including Keynesian multipliers and liquidity traps) and to Irving Fisher's debt-deflation theory.

In Chatelain and Ralf (2012), we described a research agenda for competing alternative paradigms in financial macroeconomics which should include: (1) multiple equilibria including a "persistent" systemic lack of confidence in the banking sector as modeled by Diamond and Dybvig (1983), (2) the persistent misevaluation of assets prices, which has also to take into account the probability of shifting to the systemic risk equilibrium, (3) handling clearly the contradiction between the credibility of central banks for inflation targeting versus financial stability policies such as quantitative easing during the management of systemic crisis (the literature of macro-prudential policy mostly focuses of ex ante designs assuming then that no crisis would occur) and (4) the political economy of competing jurisdictions and competing financial regulation providing loopholes for international banks, leading to banking sectors which are "fragile by political design", with an endogenously determined financial regulation which determines the probability of systemic default (Calomiris and Haber (2013)). One may try to assess the potential of Goodhart and Tsomocos framework in the current contest for paradigm shifts in financial macroeconomics.

---

[1] Centre d'Economie de la Sorbonne, Université Paris 1 Panthéon-Sorbonne, Paris School of Economics, MSE, 106-112 Boulevard de l'Hôpital, 75647 Paris Cedex 13.
[2] 368 pp. Edward Elgar Publishing, Cheltenham, Northampton, 2012, Hardback $ 160.00.

The book "*The Challenge of Financial Stability: A New Model and its Applications*" is a companion book to "*Financial Stability in Practice*", by the same authors, also published in 2012. It collects twelve articles published in peer reviewed journals from 2003 to 2010 and a computer program by K. James (section 12 of the book), along with a short introduction and a brief conclusion. Part 1 (overview) includes 2 papers. Part 2 (theory) includes 4 papers. Part 3 (applications) includes 4 papers and a computer program and finally part 4 (liquidity and collateral) includes 2 papers. The breakdown of authorship by papers is as follows: eight of them are written by C.A.E. Goodhart and D.P. Tsomocos along with different co-authors, three of them are written by D.P. Tsomocos alone or with different co-authors and a paper by Saade, Osiori and Estrada (2007) is an application of their model to the Colombian banking sector. C.A.E. Goodhart is Emeritus Professor of Banking and Finance, longstanding member of the Financial Market Group at London School of Economics. He is a world leading expert in financial stability from theoretical, applied, historical and practical point of views. Dimitrios Tsomocos is University Reader in Financial Economics and Fellow in Management in Saïd Business School and St Edmund Hall at the University of Oxford. He is a world leading expert in general equilibrium models with incomplete asset markets, money and endogenous default, following seminal works by Martin Shubik. Both authors had a long standing connection with the Bank of England.

In part 1, the first two articles provide a good overview of the research project. In particular, the second paper (Bardsen, Lindquist and Tsomocos (2008)) lists ten constraints of their research agenda: contagion, default, missing financial markets, money banks and liquidity risk, the heterogeneity of banks, macroeconomic conditions, micro-foundations, tractability, forecasting and policy analysis, testing device. Note that the internal and external political economy issues determining the framework for financial regulation are not explicitly taken into account (point 4 of the Chatelain and Ralf's research agenda), but they are tackled by C.A.E. Goodhart using alternative frameworks. The paper then compares six alternative models: real business cycles, dynamic stochastic general equilibrium (DSGE) with the financial accelerator, overlapping generation models, dynamic aggregative estimated model (Cowles commission type of models), structural vector auto-regressive, and the framework presented in the book under review: finite horizon general equilibrium models with heterogeneous agents, endogenous defaults and liquidity constraints. A few alternative paradigms are missing such as agent based simulation models.

Part 2 models theoretically the general cases for a given number of banks and a given number of households (chapter 4 and 5). Chapter 6 highlights the equivalence between a general equilibrium model with incomplete markets with endogenous default and a model with exogenous probabilities of default. Chapter 7 extends the model from two to three periods.

As part 2 may present a technical barrier to some readers not trained in general equilibrium modeling, we would recommend readers trained in theoretical or applied macroeconomics or in the microeconomics of banking to begin with part 3, in particular chapter 8, which neatly presents the three key assumptions of the core model:

(1) The model is an inter-temporal (with 2 periods) exchange economy without production, where agents reallocate endowments between date 1 and date 2 in order to smooth their consumption on both dates. Depending on the gap between their given endowments on period 1 with respect to period 2, agents are either net lenders or net borrowers.

(2) For each state of nature, borrowers choose an optimal level of strategic partial default taking into account a tradeoff where the marginal gain of default is equal to its marginal cost. This marginal cost is related to a deadweight loss, so that credit markets are imperfect. The banks and households utility are quadratic and concave: the quadratic utility function of banks' profits takes into account risk aversion. Additionally, a related (but distinct) tradeoff occurs for banks when they decide not to fulfill their regulatory equity/weighted assets ratio: they pay a penalty defined by the regulatory authorities. Credit markets are competitive in the sense that the intersections of credit demands with credit supplies determine the equilibrium. Formally, there is no asymmetric information, but moral hazard due to strategic default is dealt with by penalties, using a rational fraud (here strategic default) and punishment model à la Gary Becker. Goodhart and Tsomocos do not use incentive compatibility constraints which have the drawback to eliminate ex-ante fraud and strategic default.

(3) Contagion is modeled by a credit chain leading to a domino effect only due to the inter-bank lending market. A household ($\varphi$) is a lender and deposits in two banks and does not hold cash, so that there is not depositor's bank run. A bank ($\delta$) is a net lender to another bank ($\gamma$) which is a net borrower on the interbank market. Each bank lend to its specific nature given household (bank $\gamma$ to borrower $\alpha$, bank $\delta$ to borrower $\beta$). Borrowing households $\alpha$ or $\beta$ do not trade credit with each other or with the other bank. Hence, when household $\alpha$ increases the size of his partial default, it implies an increase of the partial default rate of bank $\gamma$ (a direct loss for depositor $\varphi$) which leads bank $\delta$ to increase its own partial default rate (an indirect loss for depositor $\varphi$). A central bank does open market operations on the interbank market where it sets the interest rate and may add liquidity. The central bank is able to limit the increase of partial default due to contagion on the inter-bank market.

At this stage, heterogeneity could be confined to be a borrower or a lender, and the characteristics of banks and households (preferences, endowments, equity, penalties) could be identical: there will be a contagion of partial default rates because of these joined hypotheses: strategic partial default and a credit chain on the interbank market.

Additional theoretical results are provided in the following chapters. Chapter 9 determines under which condition on banks side each borrowing household remains faithful to a single bank (without resorting to the assumption of switching costs for borrowers). Chapter 10 shows that an infinite horizon version of the model may be consistent with a finite horizon version. Chapter 13 extends the model with housing and mortgages. Chapter 14 is an elegant theoretical model explaining why the yield curve may be upward sloping when there is endogenous default, due to supply of liquidity by the central bank. There are no commercial banks in this last model, only a lender, a borrower and a central bank.

Before dealing with the empirical applications presented in chapters 9, 10 and 11, let us list a few potential theoretical extensions that could be added to this framework of strategic partial default with credit chain on the inter-bank market in an exchange economy.

(1) Production: One may add the growth and fluctuations of output, possibly driven by the bank credit channel of endogenous defaults. This is taken into account in the empirical application in partial equilibrium presented in chapters 9, 10 and 11.

(2) Multiple equilibria: The authors show that their model has a local unique equilibrium so that multiple equilibria could easily be taken into account. Their ex ante equilibrium takes into account two regimes for defaults on the second period: a good state with low partial defaults and a bad state (with probability suggested to by 5%) with high partial default, with large contagion effects, close to systemic default. A rational expectation ex ante equilibrium à la Diamond Dybvig (1983) with bank runs, with homogeneity of expectations of depositors and/or banks on the interbank market may also be included.

(3) Bubbles on asset prices: There may be systematic errors in the valuation of assets or in probabilities (Evanoff, Kaufman and Malliaris (2012)). To this end, one may introduce asymmetric information for investors. Or one may modify the expected utility framework where subjective and objective probabilities are identical to prospect theory or alternative non expected utility theory. Loans may be backed by the expected values of assets used as collateral, with systematic errors in its valuation.

(4) Imperfect competition: Although the number of banks is very small in the applied part 3, imperfect competition issues related to oligopolistic or local monopoly behavior of banks could be taken into account.

Those extensions would be required only if they provide additional insights to specific issues on financial stability leading to policy advice. For example, one may imagine an extension of the Goodhart and Tsomocos framework dealing with the issue of separating retail banking from investment banks (the Vickers or Glass Steagall act issue). Then, this extension would focus on assets traded on financial markets in banks' balance sheet and the two way potential contagion with credit default.

Introducing complete heterogeneity between agents adds value when applying this model in partial equilibrium for an empirical validation on a small number of large banks or of aggregated groups of banks of a given country. Indeed, each bank has a distinct size and distinct balance sheet items. This is done in chapters 9, 10 and 11 in part 3 of the book. Production is added in the model as well as the elasticities of credit and deposit with respect to output and interest rates, in order to determine the fluctuations of credit demand and deposit supply over time.

The Goodhart and Tsomocos framework has a great potential for future empirical research because of its ability to model the different items of each banks' balance sheet. Structural estimations of simultaneous equations for deposits, credit and interbank net credit for each bank may allow an identification of deep parameters in the Goodhart and Tsomocos framework, for example, risk aversion and default penalties for each banks. This will answer one of the current issues of the empirical model, which is over-fitting. The banking sector is described by too many parameters with respect to degrees of freedom in the data. Some parameters may be substitutes to others with similar effects, which is partly a theoretical and an empirical issue.

Over-fitting (too many parameters) is a problem shared by other frameworks competing with the Goodhart and Tosmocos one, such as DSGE models including the financial accelerator. When there is a competition between new paradigms, Kuhn (1962) emphasized that researchers' conversion to a new paradigm depends on a bet on the potential for solving a larger scope of problems that this new paradigm may have. This potential may offset the creative destruction cost of the former paradigm, as some phenomena will not be explained by the new paradigm. Without doubt, the Goodhart and Tsomocos framework of endogenous default with credit chain is part of a new paradigm for explaining the current financial crisis which is worth betting on it.